\documentclass[a4paper,11pt,english]{article}

\usepackage[utf8]{inputenc}
\usepackage{newtxtext} 
\usepackage[scaled=.95]{cabin} 
\usepackage[varqu,varl]{inconsolata} 
\usepackage{amsmath}
\usepackage[varg]{newtxmath}

\usepackage{amsmath}
\usepackage{graphicx}
\graphicspath{{./images/}}
\usepackage{tabularx,multirow,rotating,subfigure}
\usepackage{natbib}
\makeatletter
\newcommand\BIBand{\ifNAT@swa{\&}\else{and}\fi}
\makeatother
\usepackage{alltt}
\usepackage{tocbibind}
\usepackage{afterpage}
\usepackage{calc,ifthen}

\usepackage{url}
\usepackage{fancyheadings}

\usepackage[dvipsnames]{xcolor}
\usepackage[bookmarks,breaklinks,colorlinks,linkcolor=black,citecolor=MidnightBlue,filecolor=black,urlcolor=MidnightBlue,urlbordercolor=white]{hyperref}
\hypersetup{%
  bookmarksopen=true,
  bookmarksnumbered=true,
  pdftitle={On Reparameterization Invariant Bayesian Point
    Estimates and Credible Regions},
  pdfsubject={},
  pdfauthor={Aki Vehtari},
  pdfkeywords={Bayesian inference, point estimation, credible interval,
intrinsic estimation, predictive criterion, model selection},
  pdfstartview={FitH -32768},
  pdfborder={0 0 0},
}

\DeclareMathOperator{\E}{E}

\DeclareMathOperator{\Beta}{Beta}
\DeclareMathOperator{\Ga}{Ga}
\DeclareMathOperator{\Gg}{Gg}
\DeclareMathOperator{\Ex}{Ex}
\DeclareMathOperator{\Un}{Un}
\DeclareMathOperator{\Pa}{Pa}

\DeclareMathOperator{\Invchi2}{Inv-\chi^2}
\DeclareMathOperator{\N}{N}

\DeclareMathOperator{\KL}{\mathit{k}}

\begin{document}
\pagestyle{fancy}
\pagenumbering{arabic}
\cfoot{}
\lhead{\fancyplain{}{{\em On Bayesian Point Estimates and Credible Intervals}}}
\rhead{\fancyplain{}{\thepage}}

\addcontentsline{toc}{section}{\abstractname}

\title{\vspace{-0.4cm} On Reparameterization Invariant Bayesian\\ Point Estimates and Credible Regions}

\author{Aki Vehtari\\
  Department of Computer science\\
  Aalto University\\
  {\it Aki.Vehtari@aalto.fi}}

\maketitle

\thispagestyle{fancy}
\lhead{} \rhead{} \cfoot{}
\chead{}

\begin{abstract}  This paper considers reparameterization invariant Bayesian point
  estimates and credible regions of model parameters for scientific
  inference and communication.  The effect of intrinsic loss function
  choice in Bayesian intrinsic estimates and regions is studied with
  the following findings. A particular intrinsic loss function, using
  Kullback-Leibler divergence from the full model to the restricted
  model, has strong connection to a Bayesian predictive criterion,
  which produces point estimates with the best predictive performance.  An
  alternative intrinsic loss function, using Kullback-Leibler
  divergence from the restricted model to the full model, produces
  estimates with interesting frequency properties for at least some
  commonly used distributions, that is, unbiased minimum variance
  estimates of the location and scale parameters.
  %
%
\end{abstract}

\bigskip

Keywords: Bayesian inference, point estimation, credible interval,
intrinsic estimation, predictive criterion, model selection

\afterpage{
\chead{}
\lhead{\fancyplain{}{{\em On Reparameterization Invariant Bayesian
      Point Estimates and Credible Regions}}}
\rhead{\fancyplain{}{\thepage}}}

\section{Introduction}
\label{sec:introduction}

In the Bayesian approach, the optimal way of making inference is to
describe all the uncertainties with probabilities and probability
distributions and obtain the posterior distribution of the quantity
of interest by marginalizing over all other unknowns.
If there is uncertainty in the model structure, the optimal
Bayesian approach is to integrate also over the model space
(considering models as discrete parameters).
After taking into account everything we can think of, we have the
full (encompassing) model $M_*$ describing optimally our knowledge
of the phenomenon \citep[see, e.g.,][]{OHagan+Forster:2004,Vehtari+Ojanen:2012}.
Adequacy of the full model $M_*$ should be assessed, for example,
using external validation with new data, simulating external
validation via cross-validation
\citep[e.g.,][]{Vehtari+Ojanen:2012} or using various posterior
consistency checks
(see, e.g., \citealp{Dey+Gelfand+Swartz+Vlachos:1994}; \citealp{Gelman+Meng+Stern:1996}; \citealp[][Ch. 6]{Gelman+etal+BDA3:2013}). 

Sometimes it may be helpful to use point estimates for some of the
quantities, for example, to reduce future computational load, or as
a necessary part of reporting results. Related problems are
estimation of credible regions (or intervals in one dimension),
hypothesis testing and model selection. Credible regions are
natural extension of point estimates. Hypothesis testing can be
used to test whether a particular unknown could be fixed to a
particular value, for example, whether a covariate related
parameter could be fixed to zero. Model selection may be thought as
point estimation in the model space. 

In view of having the full model, all these problems can be
considered as model reduction problems, in which there is some
desire for using a reduced restricted model $M_R$ instead of the full
model $M_*$.
Model reduction can be considered as a decision problem, in which
there is some utility associated to the performance of
the model or loss associated to the divergence from the full model.
There may also be associated measurement and computation costs.
Preferably one should use an application specific utility (or loss
or cost) function. Sometimes this utility is not readily available
and it would then be useful to be able to use a generic reference
utility. See \citet{Vehtari+Ojanen:2012} for further discussion and references.

This paper concentrates on the point and credible region estimation
for pure scientific inference and communication using generic
reference utilities.  Connection to hypothesis testing and model
selection is discussed in the end.
Specifically the purpose of the paper is to compare the properties
of the estimates and credible regions obtained using the following
methods.
\begin{list}{$\bullet$}{}
\item Intrinsic estimation with the symmetric intrinsic loss
  function as proposed by \citet{Bernardo+Juarez:2003} and
  \citet{Bernardo:2005c}.
\item Intrinsic estimation with Kullback-Leibler divergence
  from the full model to the restricted model.
  \citet{Bernardo:1999a} used this loss function in Bayesian
  Reference Criterion before it was replaced with symmetric version
  by \citet{Bernardo+Rueda:2002}.
\item Intrinsic estimation with Kullback-Leibler divergence
  from the restricted model to the full model.
  This loss function is part of the symmetric version, but it has not
  been previously used alone.
\item Predictive criterion. This is based on a predictive model
  selection criterion proposed by
  \citet{SanMartini+Spezzaferri:1984}, which has not been
  previously used for point or credible region estimation. 
\end{list}
This paper shows a connection, including equality in
special case, between intrinsic estimation with Kullback-Leibler
divergence from the full model to the restricted model and the
predictive criterion. Examples demonstrate that the intrinsic
estimation with Kullback-Leibler divergence from the restricted
model to the full model has interesting frequency properties, that
is, it produces unbiased minimum variance estimates for certain
models with commonly used parameterization. Furthermore, it is
illustrated that the symmetric intrinsic loss function advocated by
\citet{Bernardo+Rueda:2002}, \citet{Bernardo+Juarez:2003} and
\citet{Bernardo:2005a,Bernardo:2005c} is a compromise between these
two.

To illustrate the interesting properties, this paper focuses on simple models for which we can compute the solutions analytically or with low dimensional quadrature. The predictive criterion approach is closely connected to the projective predictive approach which replaces the exact criterion with an approximation that is computationally faster \citep{Piironen+Paasiniemi+Vehtari:2020:projpred,Catalina+Buerkner+Vehtari:2020:GAMM_projpred,Catalina+Buerkener+Vehtari:2021:latent_projpred}.

Section~\ref{sec:point-estimation} reviews the point estimation as
a decision problem and preferable properties for a generic
utility (loss) function for pure scientific inference and
communication.
Section~\ref{sec:predictive-estimation} reviews the predictive
model selection criterion, shows how it can be used in point
estimation and builds up for the connection to intrinsic
estimation.
Section~\ref{sec:intrinsic-estimation} reviews intrinsic estimation
and different intrinsic loss functions used in this paper, shows
the connection of intrinsic estimation with a particular loss
function to the predictive criterion, and discusses properties of
other loss functions used in this paper.
Section \ref{sec:credible-regions} reviews credible regions, and
show how point estimation methods discussed in this paper are extended
to estimation of credible regions.
Section~\ref{sec:examples} illustrates the properties of the
methods in several examples. Derivations of the estimates are
given. In the simplest examples, the point estimates can be solved
analytically and for the rest solutions require partial numerical
computation. Numerical examples are used for further illustration.
Conclusion of the results and additional discussion are presented
in Sections \ref{sec:conclusion} and \ref{sec:discussion}.

\section{Point estimation}
\label{sec:point-estimation}

In the Bayesian approach, the uncertainties are presented with
distributions. To fully describe the information presented by the
distribution the whole distribution has to be presented \citep[see,
e.g.,][Ch. 2 and 3]{Bernardo+Smith:1994}. Point estimates may be
used to summarize the distributions for simpler communication, they
may be necessary if the action in decision problem requires stating
a single value (e.g., in control), or they may be used to simplify
the model.

Loss function can be used to measure the consequences of using a
point estimate instead of the full distribution and the point
estimate is then obtained using the decision theory.

We follow the notation by \citet{Vehtari+Ojanen:2012}. Let
$y_{(1:n)} = (y_1, \ldots, y_n)$ denote the observed data, $\tilde{y}$
denote the not yet observed future observation. We assume here the
$\mathcal{M}$-completed view
\citep{Bernardo+Smith:1994,Vehtari+Ojanen:2012}, and assume that the
posterior predictive distribution using the full model $M_*$
\begin{align}
p(\tilde{y}|y_{(1:n)},M_*) = \int p(\tilde{y}|\theta,M_*)p(\theta|y_{(1:n)},M_*) d\theta
\end{align}
is our best description of the unknown true distribution $p_t(y)$
\citep[see][for further discussion on this
assumption]{Vehtari+Ojanen:2012}.

In some cases, we may want to summarise some of the parameters
$\theta$ by point estimates.
The most commonly used Bayesian point estimates for $\theta$ are
the posterior mean, median, and mode (see, e.g., \citealp[][Ch.
5]{Bernardo+Smith:1994}; \citealp[][Ch. 4]{Robert:2001};
\citealp[][Ch. 2]{Gelman+etal+BDA3:2013}; \citealp[][Ch.
8]{Press:2003}). These correspond to using squared, absolute and
zero-one loss functions respectively and the expectation is taken
over the marginal posterior of $\theta$ given the observations
$\mathbf{y}$
\begin{equation}
  \hat{\theta}=\arg\min_{\tilde{\theta}} \int
  l(\theta,\tilde{\theta}) p(\theta|y_{(1:n)},M_*) d\theta.
\end{equation}
In some cases these loss functions may be justified by the
application, but most often they are simply used to summarize the
location of the marginal distribution.
Problem is that posterior mean and mode are not invariant under
reparameterization and median is not easily generalizable to more
than one dimension.

We also consider cases where we form point estimates only for subset
of parameters.  Let $p(y|\theta,\lambda,M_*)$ be the full model with
additional parameters $\lambda$,
and $p(y|\hat{\theta},\lambda,M_R)$ be the restricted model in which
$\theta$ is estimated with $\hat{\theta}$. Now there are a few
alternatives how to define the decision problem.
For clarity, in this paper models conditional on covariates
$p(y|x,\theta,\lambda,M)$ are not considered, since not having
explicit model for the distribution of $x$ complicates the analysis
\citep[see][for related discussion]{Vehtari+Ojanen:2012}. Such
analysis will be presented in a forthcoming paper.

\section{Loss functions}
\label{sec:point-estimation}

There are several well justified desirable properties for loss
functions in pure scientific inference. \citet{Bernardo:1979} and
\citet[][Sec. 2.7 and 3.4]{Bernardo+Smith:1994} argue that the loss
function should be proper local score function.
\citet{Robert:1996b} argues that loss function should be invariant
to reparameterization. \citet{Bernardo+Juarez:2003} elaborate that
invariance to reparameterization is necessary since in a purely
inferential context, the loss function should not measure the
discrepancy between parameter values, but directly measure the
discrepancy between the models they label. A loss function
comparing models directly $l(p,\hat{p})$ is called an intrinsic
loss \citep{Robert:1996b}. Not all intrinsic loss functions are
equally well suited, and \citet{Bernardo:2005c} gives additional
preferred condition of invariancy under reduction of the data to
sufficient statistics.
See \citet{Robert:1996b}, and \citet{Bernardo:2005a,Bernardo:2005c}
for further discussion and references.

\citet{Robert:1996b}, \citet{Bernardo+Juarez:2003}, and
\citet{Bernardo:2005a,Bernardo:2005c} consider parameterization
invariant loss functions based on comparing models
$p(\tilde{y}|\theta,\lambda,M_*)$ and $p(\tilde{y}|\theta,\lambda,M_R)$.
However, parameterization invariant loss functions can also be
achieved by comparing predictive distributions of the models
$p(\tilde{y}|y_{(1:n)},M_*)$ and $p(\tilde{y}|y_{(1:n)},M_R)$. 
The predictive model selection criterion based point estimation is
introduced in Section \ref{sec:predictive-estimation} and intrinsic
estimation, based on intrinsic loss functions for models, is
reviewed in Section~\ref{sec:intrinsic-estimation}. Extension of
point estimates to credible regions is reviewed in Section
\ref{sec:credible-regions}.

\section{Predictive criterion estimation}
\label{sec:predictive-estimation}

In prediction problems, the question in model selection is whether
some of the unknowns in the model $M_*$ could be fixed without
losing information in predictive inference. The proposed problem
may formally be described as a decision problem with two
alternating actions. One action is to predict with the full model
$M_*$ and alternative action is to use a restricted model $M_R$
with some parameters fixed. Point estimation can then be considered
as a decision problem where values for fixed parameters are chosen to
maximize the expected utility (or minimize the expected loss). See \citet{Vehtari+Ojanen:2012} for further discussion.

Let $u(\hat{p}(\tilde{y}),\tilde{y})$ be a utility function associated
with the choice of $\hat{p}(\tilde{y})$ as the predictive distribution
of the future distribution, where $\tilde{y}$ is the true unknown
future value.  The true expected utility is computed by integrating
over true distribution $p_t(\tilde{y})$
\begin{equation}
  \int u(\hat{p}(\tilde{y}),\tilde{y}) p_t(\tilde{y}) d\tilde{y},
  \label{eq:expected-utility}
\end{equation}
where $p_t(\tilde{y})$ describes the uncertainty in future value of
$\tilde{y}$.  Since the full model $M_*$ describes our knowledge of
the phenomenon, it is natural that $p(\tilde{y})$ is replaced with the
predictive distribution of the full model
$p(\tilde{y}|y_{(1:n)},M_*)$.
Thus, the expected predictive performance of the restricted model
is estimated using the full model $M_*$ as approximate ``true''
belief model.

Early account of using this approach is described in an article by
\citet{Lindley:1968a}, in which the goal was covariate selection.
The analysis was made using a Gaussian linear model and quadratic
loss function, which facilitates analytic solution. Instead of
point estimation, part of the parameters in a restricted model were
fixed to zero (thus removing their effect in the model).

\citet{Bernardo:1979} and \citet[][Ch.  2 and
3]{Bernardo+Smith:1994} argued that in a pure scientific inference
and communication context it is most appropriate to use a
logarithmic score function, which among other good properties is
also related to the information theoretic measure of the
information in the distribution. Expected utility is then
\begin{equation}
  \int \log \hat{p}(\tilde{y}) p(\tilde{y}|y_{(1:n)},M_*) d\tilde{y},
  \label{eq:expected-utility}
\end{equation}
which was called \emph{predictive model selection criterion} by
\citet{SanMartini+Spezzaferri:1984}. To take into account the point
estimation, credible regions, and hypothesis testing, we call this
simply a \emph{predictive criterion}. The predictive criterion for a
model $M_R$ is defined as
\begin{equation}
  \int \log p(\tilde{y}|y_{(1:n)},M_R) p(\tilde{y}|y_{(1:n)},M_*) d\tilde{y}.
  \label{eq:predictive-criterion}
\end{equation}
The optimal values for the fixed unknowns are obtained by
maximizing this predictive criterion and the predictive point
estimate of $\theta$ is obtained simply as
\begin{equation}
  \hat{\theta}_{\mathrm{P}}=\arg\max_{\tilde{\theta}} \int \log
  p(\tilde{y}|y_{(1:n)},\tilde{\theta},M_R) p(\tilde{y}|y_{(1:n)},M_*) d\tilde{y}.
  \label{eq:predictive-estimator}
\end{equation}
This estimate provides the best predictive distribution given the
restriction that $\theta$ is fixed instead of integrating over it.

The predictive criterion estimation approach can be considered as a
comparison of the predictive distribution of the restricted model
$M_R$ to future observations generated by the belief model $M_*$.
This is obvious if one considers the Monte Carlo approximation of
\eqref{eq:predictive-criterion},
\begin{equation}
  \frac{1}{L}\sum_{l=1}^L \log p(y^{\mathrm{rep}}_l|y_{(1:n)},M_R),
  \label{eq:predictive-estimation-mc}
\end{equation}
where $y^{\mathrm{rep}}_l$ are samples from the predictive
distribution $p(y|y_{(1:n)},M_*)$, which can be considered as 
proxy for future observations. 

For simpler models (e.g. one-parameter distributions in exponential
family) these estimates can be evaluated analytically. For more
complex models, generic approach is to use Monte Carlo sampling for
the posterior distribution and then the predictive distribution can
be approximated as a mixture distribution with components having
parameter values from the posterior samples. The Monte Carlo
approximation \eqref{eq:predictive-estimation-mc} can generally be
improved, for example, via partial use of analytic or numerical
integration. Section \ref{sec:examples} shows examples for both
simpler and more complex models.
For more complex models a fast approximate projection predictive approach can be used \citep{Piironen+Paasiniemi+Vehtari:2020:projpred,Catalina+Buerkner+Vehtari:2020:GAMM_projpred,Catalina+Buerkener+Vehtari:2021:latent_projpred}.

A usefull loss function view of the approach is obtained if the
predictive criterion of the full model $M_*$ is computed
\begin{equation}
  \int \log p(y|y_{(1:n)},M_*) p(y|y_{(1:n)},M_*) dy,
  \label{eq:entropy_of_full}
\end{equation}
and \eqref{eq:predictive-criterion} is subtracted from that to get
\begin{equation}
 \int p(y|y_{(1:n)},M_*) \log \frac{p(y|y_{(1:n)},M_*)}{p(y|y_{(1:n)},M_R)},
 \label{eq:predictive-kl}
\end{equation}
which is the Kullback-Leibler divergence from the predictive
density of the full model $p(y|y_{(1:n)},M_*)$ to the predictive
density of a restricted model $p(y|y_{(1:n)},M_R)$. Thus the use
of the predictive criterion corresponds to minimizing the loss
related to consequence of using $M_R$ instead of $M_*$ and
\begin{equation}
  \hat{\theta}=\arg\min_{\tilde{\theta}} \int
  \KL[p(y|y_{(1:n)},M_*),p(y|y_{(1:n)},M_R)] p(y|y_{(1:n)},M_*) dy,
  \label{eq:predictive-kl-est}
\end{equation}
where $\KL[p(y|y_{(1:n)},M_*),p(y|y_{(1:n)},M_R)]$ is the
Kullback-Leibler divergence \eqref{eq:predictive-kl}. From
the information theoretic point of view, the best point estimate for
the fixed unknown is such that minimum amount of information is
lost in the predictive distribution when replacing the full model
with the restricted model. The amount of information lost could be
used in hypothesis testing to make decision whether restricted
model $M_R$ can be accepted to be used instead of the full model
$M_*$. See \citet{Bernardo:1979} and \citet[][Ch. 2 and
3]{Bernardo+Smith:1994} for discussion about value of information,
and \citet{Bernardo:1999a} and \citet{Bernardo+Rueda:2002} for
discussion about calibration of that value for hypothesis testing.
The predictive distribution is naturally associated to a next
single observation in time. Alternatively, we can consider
prediction where all $n$ measurements were to be repeated again,
which adds multiplication by $n$ to logarithmic score used. The
calibration of this utility is then related to the calibration of
traditionally used likelihood-ratio. For notational and
computational convenience, in this paper the computations have been
made without multiplication by $n$, specially as it does not have
any effect on point estimates and intervals.

\citet{Laud+Ibrahim:1995} proposed 
minimization of the symmetric Kullback-Leibler divergence
\begin{equation}
\KL[p(y|y_{(1:n)},M_*),p(y|y_{(1:n)},M_R)]+\KL[p(y|y_{(1:n)},M_R),p(y|y_{(1:n)},M_*)],
\end{equation}
where the second term is the Kullback-Leibler divergence from the
predictive density of a restricted model $p(y|y_{(1:n)},M_R)$ to
the predictive density of the full model $p(y|y_{(1:n)},M_*)$.
The first term has a predictive interpretation since it differs
from \eqref{eq:predictive-criterion} just by a constant
\eqref{eq:entropy_of_full}. The second term does not have this
interpretation since entropy is not constant when optimizing the
parameters of the restricted model. Thus sum of these terms does
not have a predictive interpretation.

There are also other approaches
\citep[e.g.,][]{Laud+Ibrahim:1995,Gelfand+Ghosh:1998,Gutierrez-Pena+Walker:2001}
which are predictive approaches, but which are not based on
estimating the expected predictive performance of the restricted
model using the full model $M_*$ as an approximate true belief
model.

\section{Intrinsic estimation}
\label{sec:intrinsic-estimation}

Intrinsic estimation is a likelihood-based method, where the focus
is in the likelihood instead of the predictive distribution.
Intrinsic estimation is related to the Bayesian reference criterion
\citep{Bernardo:1999a,Bernardo+Rueda:2002}. \citet{Bernardo:1999a}
proposed to use a directed Kullback-Leibler divergence, while
\citet{Bernardo+Rueda:2002} and \citet{Bernardo+Juarez:2003}
proposed to use a symmetric Kullback-Leibler divergence.
\citet{Bernardo:2005a,Bernardo:2005c} provides additional
discussion and examples on intrinsic estimation and regions.

Intrinsic estimation by \citet{Bernardo+Juarez:2003} is based
on the intrinsic discrepancy, which belongs to the class of
intrinsic loss functions \citep{Robert:1996b}.  The intrinsic
discrepancy $\delta\{p_1,p_2\}$ between two distributions is
defined by \citet{Bernardo+Rueda:2002} and
\citet{Bernardo+Juarez:2003} as
\begin{equation}
  \delta\{p_1,p_2\}=\min\left\{\int
    p_1(y)\log\frac{p_1(y)}{p_2(y)}dy,\int
    p_2(y)\log\frac{p_2(y)}{p_1(y)}dy \right\}, 
  \label{eq:intrinsic-discrepancy}
\end{equation}
which is the minimum of two Kullback-Leibler divergences: one
from $p_1$ to $p_2$ and another from $p_2$ to $p_1$.
The intrinsic discrepancy $\delta\{M_1,M_2\}$ between two models is
defined as the minimum intrinsic discrepancy between their elements
\begin{equation}
  \delta\{M_1,M_2\} =
  \inf_{\theta_1,\theta_2}\delta\left\{p(y|\theta_1,M_1),p(y|\theta_2,M_2)\right\}.
  \label{eq:intrinsic-discrepancy}
\end{equation}
For intrinsic estimation consider the full model
$p(y|\theta,\lambda,M_*)$ and a restricted model
$p(y|\tilde{\theta},\lambda,M_R)$, where $\lambda$ may be
considered as nuisance parameter.
The intrinsic estimator is obtained by minimizing the reference
posterior expectation of the intrinsic discrepancy
\begin{equation}
  \hat{\theta}_{\mathrm{I}}=\arg\min_{\tilde{\theta}} \iint
  \min_{\tilde{\lambda}}\delta\left\{p(y|\tilde{\theta},\tilde{\lambda},M_R),p(y|\theta,\lambda,M_*)\right\} p(\theta,\lambda|y_{(1:n)},M_*) d\theta d\lambda.
  \label{eq:intrinsic-estimator}
\end{equation}

If the Kullback-Leibler divergence (discrepancy) minimization view
of the predictive approach is considered (see
\eqref{eq:predictive-kl-est}), the similarity to intrinsic
estimation is obvious. In the predictive approach the unknown not
fixed parameters are integrated out for each model, and posterior
predictive distributions are compared directly, while in intrinsic
estimation the discrepancy of likelihoods is evaluated first and
the expectation is computed by integrating over the posterior of
$\theta$ and $\lambda$ given the full model $M_*$.
Estimator \eqref{eq:intrinsic-estimator} may be generally more
difficult to compute than \eqref{eq:predictive-estimator}, since
this requires minimization inside the integration, although for
simpler models this minimization is solvable analytically, but more
complex models require numerical integration.

Intrinsic estimation is not directly applicable for hyperparameters
in hierarchical models. The intrinsic discrepancy
\eqref{eq:intrinsic-discrepancy} does not have reference to
hyperparameters as the discrepancy is computed using only the
likelihood part of the model. To use intrinsic estimation, one has
to compute the marginal likelihood by integrating over all lower
level parameters, which may be difficult for complex hierarchical
models.
\citet{Trevisani+Gelfand:2003a} discuss a problem of focus in
hierarchical models when using likelihood-based methods. Intrinsic
estimation does not have this problem since the focus is determined
by which parameters are going to be estimated.

In this paper, three intrinsic discrepancy loss functions compared
are:
\begin{eqnarray}
  \delta_1(\tilde{\theta},\theta) & = & \KL(\tilde{\theta}|\theta) \\
  \delta_2(\tilde{\theta},\theta) & = & \KL(\theta|\tilde{\theta}) \\
  \delta_3(\tilde{\theta},\theta) & = & \min \{
      \KL(\tilde{\theta}|\theta),  \KL(\theta|\tilde{\theta})
      \},
\end{eqnarray}
where $\KL(\tilde{\theta}|\theta)$ and $\KL(\theta|\tilde{\theta})$
are shorthand notations for
$\KL[p(y|\tilde{\theta},M_R),p(y|\theta,M_*)]$ and
$\KL[p(y|\theta,M_*),p(y|\tilde{\theta},M_R)]$ respectively.
$\delta_1$ was used by \citet{Bernardo:1999a}, $\delta_3$ by
\citet{Bernardo+Rueda:2002} and \citet{Bernardo+Juarez:2003}, and
$\delta_2$ has not been used before, but it has some interesting
frequency properties as illustrated later.
$\delta_1$ was also used by \citet{Goutis+Robert:1998a} and
\citet{Dupuis+Robert:2003}, but using approximate Kullback-Leibler
projections of the full model parameters to a restricted parameter
space (see also later development of this approach by \citealp{Piironen+Paasiniemi+Vehtari:2020:projpred}, \citealp{Catalina+Buerkner+Vehtari:2020:GAMM_projpred}, and \citealp{Catalina+Buerkener+Vehtari:2021:latent_projpred}).

Likelihood $p(y|\cdot)$ for several observations is a form
$\prod_{i=1}^n p(y_i|\cdot)$. Discrepancies are equal for each $i$,
and thus total discrepany is $n$ times single discrepancy. Since
this scaling does not affect point and interval estimation,
multiplication by $n$ is omitted for notational and computational
convenience. For calibration of the discrepancy to the
likelihood-ratio scale multiplication by $n$ is used.

Intuitively, the intrinsic discrepancy $\delta_1$ is related to the
predictive approach, since the Kullback-Leibler divergence is
computed from the full model to the restricted model. If the point
estimate is made for all the likelihood parameters (no nuisance
parameters $\lambda$ in \eqref{eq:intrinsic-estimator}), then
relation is exact. This can be seen by rearranging the terms in
the expected intrinsic discrepancy as
\begin{equation}
  \begin{split}\raisetag{\baselineskip}
    &\iint p(y|\theta,M_*) \log \frac{p(y|\theta,M_*)}{p(y|\tilde{\theta},M_R)}dy
  p(\theta|y_{(1:n)},M_*) d\theta\\
  &= \iint p(y|\theta,M_*) \log p(y|\theta,M_*)
  p(\theta|y_{(1:n)},M_*) dy\theta
  - \iint p(y|\theta,M_*) \log p(y|\tilde{\theta},M_R)
  p(\theta|y_{(1:n)},M_*) dy\theta \\
  &= C - \iint p(y|\theta,M_*) p(\theta|y_{(1:n)},M_*) d\theta 
  \log p(y|\tilde{\theta},M_R) dy \\
  &= C - \int p(y|y_{(1:n)},M_*) \log p(y|\tilde{\theta},M_R) dy.
  \end{split}
  \label{eq:dir1-intr-equal-pred}
\end{equation}
The second term is equal to the negative predictive criterion with
$n$ predictions (see previous section). The first term is constant
$C$, which is not dependent on $\tilde{\theta}$. In the predictive
criterion the corresponding term is the predcitive criterion of the
full model $M_*$ \eqref{eq:entropy_of_full}, but since the order of
integration and logarithmic function is changed, the terms produce
different values.
Difference up to a constant does not have effect for the shape of
the loss function given $\theta$, and thus, in this special case,
both methods produce exactly the same point estimate.

If point estimate is computed only for some of the parameters, the
minimization over $\tilde{\lambda}$ in
\eqref{eq:intrinsic-estimator} makes relation more complex. In the
predictive approach the predictive distribution of the restricted
model $M_R$ is obtained simply by integrating over the posterior
distribution $p(\lambda|\tilde{\theta},y_{(1:n)},M_R)$. In the
intrinsic approach, taking the infimum of the discrepancy over
$\lambda$ (see \eqref{eq:intrinsic-estimator}) corresponds to
replacing the restricted model $M_R$ with Kullback-Leibler
projection of the the full model $M_*$. Projected model is
$p(y|\tilde{\theta},\lambda^\bot,M_R)$, where the distribution of
the $\lambda^\bot$ is the Kullback-Leibler projection of the
distribution $p(\lambda|y_{(1:n)},M_*)$.
Using projection changes the restricted model and thus the decision
problem is not exactly same in these two approaches.
Point estimates and credible regions can still be similar as
illustrated in Section \ref{sec:example-normal-mean} with normal
model with variance as a nuisance parameter.
For further discussion and examples of defining submodels as the
Kullback-Leibler projections of the full model, see papers by
\citet{Piironen+Paasiniemi+Vehtari:2020:projpred}, \citet{Catalina+Buerkner+Vehtari:2020:GAMM_projpred}, and \citet{Catalina+Buerkener+Vehtari:2021:latent_projpred}.

In the case of the intrinsic discrepancy $\delta_2$, the divergence
is computed from the restricted model to the full model. Thus, for
given values of $\theta$, it is assumed that the restricted model
is the ``true'' model.
This case differs from the case of computing the Kullback-Leibler
divergence from the predictive distribution of the restricted model
to the predictive distribution of the full model, since the
expectation is still taken over the posterior distribution of
$\theta$ given the full model $M_*$.
This approach could be interpreted as trying to find a restricted
``truth'' which optimally agrees with the posterior information of
the full model. This is related to the frequentist view of assuming
that there is a true value which generated the data.
In section \ref{sec:examples}, it is demonstrated that at least in
the case of three regular problems with specific commonly used
parameterizations, $\delta_2$ produces estimates which have
frequency properties equal to unbiased minimum variance estimates.
Equality is valid only for specific parameterizations, since
unbiased estimates are not invariant under reparameterization. 
Note that, here unbiasedness is a side effect instead of an actual
design goal.
For general criticism against unbiasedness as a design goal for
generic estimates, see \citet[][Ch. 5]{OHagan+Forster:2004}. For
advocacy of frequentist evaluation of the Bayesian methods, see,
for example, \citet{Bayarri+Berger:2004a} and
\citet{Bernardo:2005b}.

\citet{Bernardo+Rueda:2002} introduced the use of the symmetric
discrepancy $\delta_3$ \eqref{eq:intrinsic-discrepancy}. They claim
that it addresses two ``unwelcome'' features of directed
Kullback-Leibler divergence; it is not symmetric, and it diverges
if the support of $p_2(y)$ is a strict subset of the support of
$p_1(y)$.
In the predictive approach, it is obvious that the symmetric
Kullback-Leibler divergence is in conflict with the use of the full
model as an approximate ``true'' belief model for future
observations. In intrinsic estimation, this is not as clear since
the expectation of the discrepancy is still taken over the
posterior of the parameters of the full model.
Examples in Section \ref{sec:examples} demonstrate that at least in
four regular problems, the symmetric divergence produces results
which are exactly or approximately average of results obtained with
two directed divergences.
\citet{Bernardo+Rueda:2002} and \citet{Bernardo+Juarez:2003} argue
that important feature of the symmetric divergence is that it does
not diverge if the support of $p_2(y)$ is a strict subset of the
support of $p_1(y)$. \citet{Bernardo+Juarez:2003} illustrate this
with a non-regular problem, where the directed Kullback-Leibler
divergences diverge.
Although providing an estimate in non-regular problem, a convincing
justification of the symmetric divergence seems to be missing.
In Section \ref{sec:example-uniform} the non-regular problem used
by \citet{Bernardo+Juarez:2003} is reviewed and used to illustrate
additional discussion on this topic.

\section{Credible regions}
\label{sec:credible-regions}

Credible regions are natural extension of point estimates
summarizing which values are close to a point estimate and thus
describe the associated uncertainty in the posterior distribution.

Most commonly used Bayesian credible regions are the central
interval and highest posterior density (HPD) region (see, e.g.,
\citealp[][Ch. 5]{Bernardo+Smith:1994}; \citealp[][Ch.
4]{Robert:2001}; \citealp[][Ch. 2]{Gelman+Carlin+Stern+Rubin:2003};
\citealp[][Ch. 8]{Press:2003}). Central interval is defined so that
there is equal amount of probability mass outside of both ends of
the interval. This is not sensible, for example, if maximum of the
posterior is at the edge of of the parameter space or there is very
low posterior density in the center of the interval. Furthermore,
central interval is not generalizable to several dimensions. HPD
region is defined so that inside the region the posterior density
is everywhere higher than outside the region. The problem is that
HPD region is not invariant under reparameterization. For further
illustration of the problems of these two credible regions see
\citet{Bernardo:2005c}.

Decision theoretic approach for point estimates can be naturally
extended to regions.
\citet{Bernardo:2005a,Bernardo:2005c} proposed selecting credible
regions as a lowest posterior loss (LPL) regions, where all points
in the region have smaller posterior expected loss than all points
outside. These regions are form
\begin{equation}
  C_q\equiv\{\tilde{\theta};d(\tilde{\theta}|y_{(1:n)})\leq l(q)\},
\end{equation}
such that 
\begin{equation}
  \int_{C(q)} p(\theta|y_{(1:n)})d\theta =q.
\end{equation}
\citet{Bernardo:2005a,Bernardo:2005c} defines intrinsic credible
region as the lowest posterior loss region with respect to the
intrinsic discrepancy loss, and the appropriate reference prior.

In similar way, the predictive criterion credible region can be
defined as the highest posterior utility region with respect to the
predictive criterion.
Note that, if there is no nuisance parameters $\lambda$, the
intrinsic credible region with loss function
$\KL(\tilde{\theta}|\theta)$ is equal to the predictive criterion
region. This follows from the relation of the corresponding
expected intrinsic loss and the predictive criterion as discussed in
the previous section.

Intrinsic credible regions and predictive criterion credible
regions have corresponding good properties as the respective point
estimates, such as invariancy under reparameterization.
One way to evaluate credible regions is to examine their
frequentist coverage. \citet{Bernardo:2005c} shows that for certain
models all reference posterior credible regions are exact
frequentist confidence regions, that is, their frequentist coverage
is exact. Thus, for such models (with reference prior) all
intrinsic credible regions and predictive criterion credible
regions are exact confidence regions. Some of these models are
presented in Section \ref{sec:examples}. Furthermore, in Section
\ref{sec:example-bernoulli}, binomial model is used to illustrate
the frequentist coverage of credible regions in case for which
exact confidence regions generally do not exist.

\section{Examples}
\label{sec:examples}

The discussion is illustrated with five examples: Normal model with
unknown mean and variance (joint estimation and estimation under
presence of nuisance parameter), Binomial model, Exponential model,
Uniform model (non-regular model), and two-level hierarchical
Normal model.

Examples illustrate the strong relation of intrinsic estimation
with loss function $\KL(\tilde{\theta}|\theta)$ to estimation with
the predictive criterion, the relation of intrinsic estimation with
loss function $\KL(\theta|\tilde{\theta})$ to minimum variance
unbiased estimates, and that the symmetric intrinsic loss function
advocated by \citet{Bernardo+Rueda:2002},
\citet{Bernardo+Juarez:2003} and
\citet{Bernardo:2005a,Bernardo:2005c} is a compromise between these
two.

\subsection{Normal model with unknown mean and variance}
\label{sec:example-normal}

Let data $y_{(1:n)}=\{y_1,\ldots,y_n\}$ be a random sample from a
Normal distribution $\N(y|\mu,\sigma^2)$, where both $\mu$ and
$\sigma$ are unknown, and consider the problem of point estimation
of $\mu$ and $\sigma$. As a numerical illustration an example by
\citet{Bernardo:2005c} is used, with $n=25$,
$\bar{y_{(1:n)}}=0.024$, and
$\frac{1}{n}\sum_{i=1}^n(y_i-\bar{y})^2=1.077$.
The reference prior is $1/\sigma^2$ \citep{Bernardo:1979b}.

\subsubsection{Predictive criterion estimation}
The predictive distribution for the full model is
\begin{equation}
  t_{n-1}(y|\bar{y_{(1:n)}},(1+1/n)s^2), 
\end{equation}
where $s^2=\frac{1}{n-1}\sum_{i=1}^n(y_i-\bar{y})^2$.
The predictive distribution of the restricted model is
\begin{equation}
  N(y|\tilde{\mu},\tilde{\sigma}^2). 
\end{equation}
The predictive criterion to maximize is
\begin{equation}
  \int \log N(y|\tilde{\mu},\tilde{\sigma}^2)
  t_{n-1}(y|\bar{y_{(1:n)}},(1+1/n)s^2) dy,
\end{equation}
which can be efficiently evaluated given $\tilde{\mu}$ and
$\tilde{\sigma}^2$, for example, with adaptive quadrature methods.

\subsubsection{Intrinsic estimation}

For Normal model, Kullback-Leibler divergences are
\begin{eqnarray}
  \KL(\tilde{\mu},\tilde{\sigma}|\mu,\sigma)&=& \int
  \N(y|\mu,\sigma^2)\log\frac{\N(y|\mu,\sigma^2)}{\N(y|\tilde{\mu},\tilde{\sigma}^2)} dy \\
  &=& \frac{1}{2} \left[ \frac{(\mu-\tilde{\mu})^2}{\tilde{\sigma}^2} 
    + \frac{\sigma^2}{\tilde{\sigma}^2}
    + \log\left(\frac{\tilde{\sigma}^2}{\sigma^2}\right)
    - 1 
  \right],
  \label{eq:normal-dir1-discr}
\end{eqnarray}
and
\begin{eqnarray}
  \KL(\mu,\sigma|\tilde{\mu},\tilde{\sigma})&=& \int
  \N(y|\tilde{\mu},\tilde{\sigma}^2)\log\frac{\N(y|\tilde{\mu},\tilde{\sigma}^2)}{\N(y|\mu,\sigma^2)} dy \\
  &=& \frac{1}{2} \left[ \frac{(\tilde{\mu}-\mu)^2}{\sigma^2}
    + \frac{\tilde{\sigma}^2}{\sigma^2}
    + \log\left(\frac{\sigma^2}{\tilde{\sigma}^2}\right)
    -1
  \right]. 
\end{eqnarray}
The intrinsic discrepancy to minimize is
\begin{align}
  d(\tilde{\theta},y_{(1:n)}) & = \int
  \delta(\tilde{\theta},\theta)p(\theta|y_{(1:n)})d\theta,\\
  \intertext{which can be approximated using Monte Carlo} 
  d(\tilde{\theta},y_{(1:n)}) & \approx \frac{1}{T} \sum_{t=1}^T
  \delta(\tilde{\theta},\theta^{(t)}),
\end{align}
where $\theta^{(t)}$ are samples from $p(\theta|y_{(1:n)})$.

\subsubsection{Results}

Table \ref{tab:results1} shows point estimates of $\mu$ and
$\sigma$ obtained with different methods. Note that, predictive
criterion and intrinsic estimation methods are invariant under
reparameterization and thus, for example,
$\sigma^*=\sqrt{{\sigma^2}^*}$, but unbiased estimate of $\sigma^2$
is not unbiased estimate of $\sigma$. 
\begin{table}[htbp]
  \centering
  \begin{tabular}[t]{l c c}
    Approach & $\mu^*$ & $\sigma^*$ \\ \hline
    Predictive criterion estimation & 0.024 & 1.171 \\
    Intrinsic estimation directed $\KL(\tilde{\theta}|\theta)$ & 0.024 & 1.171 \\
    Intrinsic estimation directed $\KL(\theta|\tilde{\theta})$ &
    0.024 & 1.099 \\
    Intrinsic estimation symmetric $\min\{\KL(\tilde{\theta}|\theta),\KL(\theta|\tilde{\theta})\}$ & 0.024 & 1.133 \\
    Minimum variance unbiased estimate & 0.024 & 1.099 \\
  \end{tabular}
  \caption{Point estimate of $\mu$ and $\sigma$ of Normal model
    with different methods. Note that, minimum variance unbiased
    estimate is actually for $\mu,\sigma^2$ and reported estimate
    $\sqrt{{\sigma^2}^*}$ is not unbiased estimate of $\sigma$.}
  \label{tab:results1}
\end{table}

Predictive criterion estimation and intrinsic estimation with
directed divergence $\KL(\tilde{\theta}|\theta)$ give exactly the
same estimate (see section \ref{sec:intrinsic-estimation}). Both
methods assume that $\mu$ and $\sigma$ are unknown and one is
interested to find out the best single proxy values for these
parameters so that the description of the future uncertainty
(predictive distribution) changes as little as possible. Intrinsic
estimation with $\KL(\tilde{\theta}|\theta)$ can be made in
following steps
\begin{align}
\text{$\KL(\tilde{\mu},\tilde{\sigma}|\mu,\sigma)$ is minimized wrt. $\tilde{\mu}$
  setting } &
  \tilde{\mu}  =  \mu \\
  \text{integrating over the posterior gives} & 
  \mu^*  =  \E[\mu] = 0.024 \\
  \text{using above, $\KL(\tilde{\mu},\tilde{\sigma}|\mu,\sigma)$ is minimized wrt.
    $\tilde{\sigma}$ setting }  & \tilde{\sigma}^2 = \sigma^2+(\mu-\mu^*)^2 \\
  \text{integrating over posterior gives }  & 
  \sigma^* = \sqrt{\E[\sigma^2+(\mu-\mu^*)^2]} \approx 1.171,
\end{align}
that is, the estimate for $\sigma$ combines the posterior
information about $\sigma$ and posterior uncertainty about $\mu$.

The frequentist minimum variance unbiased estimate is based on the
assumption that there exists a true fixed value. Intrinsic
estimation with directed $\KL(\theta|\tilde{\theta})$ makes a
similar assumption by assuming that the true belief model is
the restricted model. These two methods give same result (up to
a Monte Carlo error).

Intrinsic estimation with the symmetric divergence gives an estimate
which is between these, and more specifically for the precision
$1/\sigma^2$ the estimate is approximately arithmetic average of
Bayesian predictive estimate and minimum variance unbiased
estimate.

\begin{figure}[htbp]
  \centering
  \includegraphics{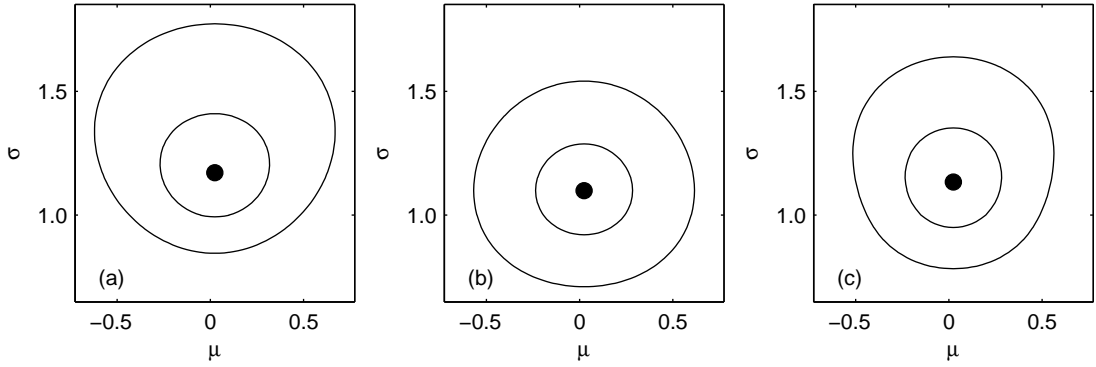}
  \caption{Point estimates (solid dot) and credible regions (50\%
    and 95\%) for $\mu$ and $\sigma$: (a) Predictive criterion
    estimation and intrinsic estimation with directed
    $\KL(\tilde{\theta}|\theta)$, (b) Intrinsic estimation directed
    $\KL(\theta|\tilde{\theta})$, (c) Intrinsic estimation
    symmetric
    $\min\{\KL(\tilde{\theta}|\theta),\KL(\theta|\tilde{\theta})\}$.}
  \label{fig:normregions}
\end{figure}

Figure \ref{fig:normregions} shows the credible regions with
different methods. Predictive criterion and intrinsic credible
regions with directed divergence $\KL(\tilde{\theta}|\theta)$ are
exactly the same and shown in same subplot. Credible regions
reflect the same properties as the respective point estimates. Note
that in this case, all these regions are exact confidence regions
\citep[see ][]{Bernardo:2005c}.

\subsubsection{Estimation under presence of nuisance parameter}
\label{sec:example-normal-mean}

Point estimation may be used also only for some of the parameters
and the rest may be considered as nuisance parameters. In such
case, the predictive criterion and the expected intrinsic
discrepancy with $\KL(\tilde{\theta}|\theta)$ are not generally
equal as in \eqref{eq:intrinsic-estimator}.
As illustration, the same Normal model example is used, except only
the mean in the restricted model is estimated and variance is
considered as a nuisance parameter.

The posterior distribution of the $\sigma^2$ is (using notation by
\citet{Gelman+Carlin+Stern+Rubin:2003})
\begin{equation}
  \Invchi2(n,\frac{1}{n}\sum_{i=1}^n(y_i-\tilde{\mu})^2),
  \label{eq:normal-muonly-sigma-post}
\end{equation}
the predictive distribution of the restricted model is
\begin{equation}
  t_{n}(y|\tilde{\mu},\frac{1}{n}\sum_{i=1}^n(y_i-\tilde{\mu})^2),
\end{equation}
and the predictive criterion to maximize is
\begin{equation}
  \int \log t_{n}(y|\tilde{\mu},\frac{1}{n}\sum_{i=1}^n(y_i-\tilde{\mu})^2)
  t_{n-1}(y|\bar{y_{(1:n)}},(1+1/n)s^2) dy.
  \label{eq:pred-discr-normal-muonly}
\end{equation}

In intrinsic estimation parameters, which are not to be estimated,
are chosen so that the intrinsic loss between models is minimized.
The directed intrinsic discrepancy \eqref{eq:normal-dir1-discr} is
minimized by setting
\begin{equation}
  \tilde{\sigma}^2=(\tilde{\mu}-\mu)^2+\sigma^2.
\end{equation}
This can be considered as Kullback-Leibler projection of $\sigma^2$
of the full model. It is easy to see that distribution of this
projection is different than \eqref{eq:normal-muonly-sigma-post}.
The intrinsic discrepancy is
\begin{equation}
  \frac{1}{2}\log\left(\frac{(\mu-\tilde{\mu})^2}{\sigma^2}+1\right),
\end{equation}
and it's exepcation can be estimated numerically.

\begin{figure}[t]
  \centering \includegraphics{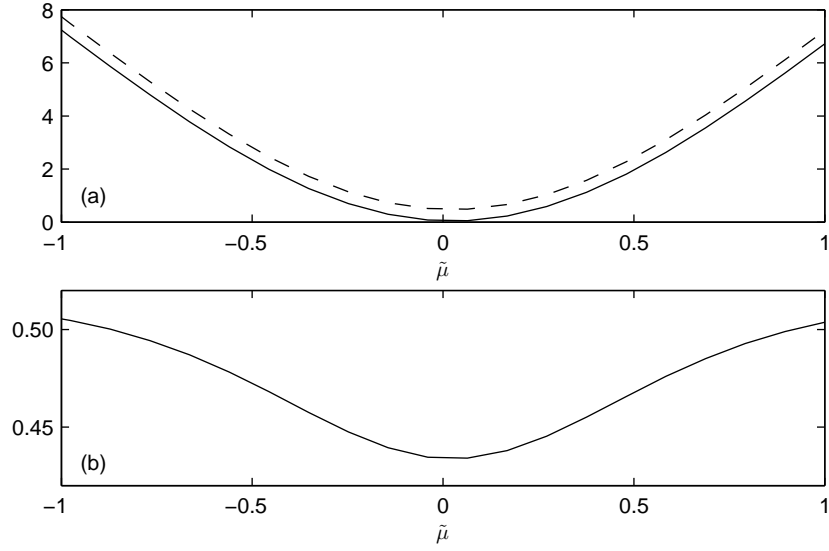}
  \caption{Top plot (a) The predictive criterion (solid line) and the
    expected intrinsic discrepancy with directed
    $\KL(\tilde{\theta}|\theta)$ (minus $C$, dashed line) for
    different values of $\tilde{\mu}$. Bottom plot (b) The
    difference of the above two discrepancies.}
  \label{fig:normmuonly}
\end{figure}

Top plot of the Figure \ref{fig:normmuonly} shows the predictive
criterion and the expected intrinsic discrepancy for different
values of $\tilde{\mu}$ (). Bottom plot of the Figure
\ref{fig:normmuonly} shows the difference of these two. For
comparison purposes the predictive criterion has been subracted
from the predictive criterion of the full model. Both discrepancies
have been scaled by $n$. Functions are very similar, where most of
the difference is due to constant term differnce (see Section
\ref{sec:intrinsic-estimation}. Despite the difference, since
functions are symmetric and minimum is in same point, the point
estimate and the LPL-interval are equal.

\subsection{Binomial model}
\label{sec:example-bernoulli}

Let $y_{(1:n)}=\{y_1,\ldots,y_n\}$, be a random sample from the
Binomial distribution with parameter $\theta$. As a numerical
illustration an example by \citet{Bernardo:2005c} is used,
with $n=10$ and $r=0$. The reference prior is
$\Beta(\frac{1}{2},\frac{1}{2})$.

\subsubsection{Predictive criterion estimation}

Predictive probability that the next observation is $1$ for the
full model is
\begin{align}
  \Pr(\tilde{y}=1|y_{(1:n)})&=\int_0^1 \Pr(\tilde{y}=1|\theta,y_{(1:n)})p(\theta|y_{(1:n)})d\theta\\
  &=\int_0^1\theta p(\theta|y_{(1:n)})d\theta=\E(\theta|y_{(1:n)})=\frac{r+1/2}{n+1},
\end{align}
and for the restricted model
\begin{align}
  \Pr(\tilde{y}=1|\tilde{\theta})&=\tilde{\theta}.
\end{align}
Taking into account the corresponding probabilities for
$\Pr(\tilde{y}=0|\cdot)$ the predictive criterion to maximize is
\begin{equation}
    \int \log
    \Pr(\tilde{y}=y|\tilde{\theta})\Pr(\tilde{y}=y|y_{(1:n)}) dy 
    = \log \tilde{\theta} \left(\frac{r+1/2}{n+1}\right) + \log
    (1-\tilde{\theta}) \left(\frac{n-r+1/2}{n+1}\right),
  \label{eq:pred-discr-bern}
\end{equation}
which is maximized by 
\begin{equation}
  \hat{\theta}=\frac{r+1/2}{n+1},
  \label{eq:pred-est-bern}
\end{equation}
that is, the posterior mean of $\theta$.

\subsubsection{Intrinsic estimation}

Directed Kullback-Leibler divergences are
\begin{equation}
  \KL(\tilde{\theta}|\theta)=\theta\log[\theta/\tilde{\theta}]+(1-\theta)\log [(1-\theta)/(1-\tilde{\theta})],
  \label{eq:dir1-intr-discr-exp}
\end{equation}
and
\begin{equation}
  \KL(\theta|\tilde{\theta})=\tilde{\theta}\log[\tilde{\theta}/\theta]+(1-\tilde{\theta})\log [(1-\tilde{\theta})/(1-\theta)].
\end{equation}
The symmetric intrinsic discrepancy between
$p(y_{(1:n)}|\tilde{\theta})$ and $p(y_{(1:n)}|\theta)$ is
  \begin{gather}
    \delta_3(\tilde{\theta},\theta)=
    \begin{cases}
      \KL(\theta|\tilde{\theta}) & \theta \in (\tilde{\theta},1-\tilde{\theta}),\\
      \KL(\tilde{\theta}|\theta) & \text{otherwise}.
    \end{cases}
  \end{gather}
The expected posterior intrinsic discrepancy is
\begin{align}
  d(\tilde{\theta},y_{(1:n)})=\int_0^1\delta(\tilde{\theta},\theta)\Beta(\theta|r+1/2,n-r+1/2)d\theta.
\end{align}
For the directed discrepancy $\KL(\tilde{\theta}|\theta)$ this can be
easily solved by taking out the terms independent of
$\tilde{\theta}$ to obtain
\begin{equation}
  \begin{split}
    - \int_0^1
    [\theta\log(\tilde{\theta})+(1-\theta)\log(1-\tilde{\theta})]\Beta(\theta|r+1/2,n-r+1/2)d\theta \\
    = -\log \tilde{\theta} \left(\frac{r+1/2}{n+1}\right) - \log
    (1-\tilde{\theta}) \left(\frac{n-r+1/2}{n+1}\right),
\end{split}
\end{equation}
which is proportional to the negative predictive criterion
\eqref{eq:pred-discr-bern} and minimized by
\begin{equation}
  \hat{\theta}=\frac{r+1/2}{n+1}.
  \label{eq:dir1-intr-est-bern}
\end{equation}
There is analytical solution also for the directed discrepancy
$\KL(\theta|\tilde{\theta})$ given by \citet{Bernardo+Juarez:2003}
\begin{equation}
  \hat{\theta}=\frac{\exp[\psi(r+1/2)]}{\exp[\psi(r+1/2)]+\exp[\psi(n-r+1/2)]},
  \label{eq:dir2-intr-est-bern}
\end{equation}
where $\psi(\cdot)$ is the digamma function. This can be quite well
approximated with
\begin{gather}
  \hat{\theta}\approx
  \begin{cases}
    \frac{r+1/7}{n+2/7} & r=0,r=n \\
    \frac{r}{n} & \text{otherwise}.
  \end{cases}
\end{gather}
The expectation of the symmetric discrepancy can be solved
numerically. \citet{Bernardo:2005a} proposed also the following
linear approximation
\begin{equation}
  \hat{\theta}  \approx \frac{r+1/3}{n+2/3}.
  \label{eq:sym-intr-est-bern}
\end{equation}

\subsubsection{Results}

Table \ref{tab:results2} shows point estimates and credible
intervals of $\theta$ obtained with different methods.
Since data are discrete, exact confidence interval does not
generally exist, and thus credible intervals produced can only be
approximate confidence intervals. There are many proposed
approximations with different desired properties but no consensus of
which one should be preferred \citep{Brown+Cai+Dasgupta:2001a}, and
thus comparison to them is not included.
\begin{table}[htbp]
  \centering
  \begin{tabular}[t]{l c c}
    Approach & $\hat{\theta}$ & CI \\ \hline
    Predictive criterion estimation & 0.045 & (0.0,\,0.305) \\
    Intrinsic estimation directed $\KL(\tilde{\theta}|\theta)$ & 0.045  & (0.0,\,0.305) \\
    Intrinsic estimation directed $\KL(\theta|\tilde{\theta})$ & 0.014  & (0,\,0.171) \\
    Intrinsic estimation symmetric $\min\{\KL(\tilde{\theta}|\theta),\KL(\theta|\tilde{\theta})\}$ & 0.031  & (0,\,0.171) \\
    Minimum variance unbiased estimate & 0 &  \\ 
  \end{tabular}
  \caption{Point estimates and 95\%-credible intervals (CI)
    of $\theta$ of Binomial model with different methods.}
  \label{tab:results2}
\end{table}

Predictive criterion estimation and intrinsic estimation with the
directed discrepancy $\KL(\tilde{\theta}|\theta)$ give exactly the
same point estimate (see \eqref{eq:pred-est-bern} and
\eqref{eq:dir1-intr-est-bern}) and also exactly the same credible
interval.

The intrinsic estimation with the directed discrepancy
$\KL(\theta|\tilde{\theta})$ gives a much lower estimate, but not
exactly same as the minimum variance unbiased estimate, which can
be explained by the influence of the prior.
If the prior $\Beta(a,a)$ is used and $a\rightarrow 0$ intrinsic
estimate with the directed discrepancy $\KL(\theta|\tilde{\theta})$
goes to 0.
If using the reference posterior, but with $r=1$, the estimate is
$0.103$ which is quite close to the corresponding minimum variance
unbiased estimate $0.1$. In fact, the approximation
$\hat{\theta}\approx r/n$ is very good except for $r=0$ and $r=n$.

Intrinsic estimation with the symmetric divergence gives an
estimate which is between the directed estimates, and more
specifically the estimate is approximately the arithmetic average
of the directed estimates. \citet{Bernardo+Juarez:2003} use this
average as an approximation to the exact result with the symmetric
discrepancy, without commenting the relevance of the estimates
obtained with directed discrepancies.

\begin{figure}[htbp]
  \centering
  \includegraphics{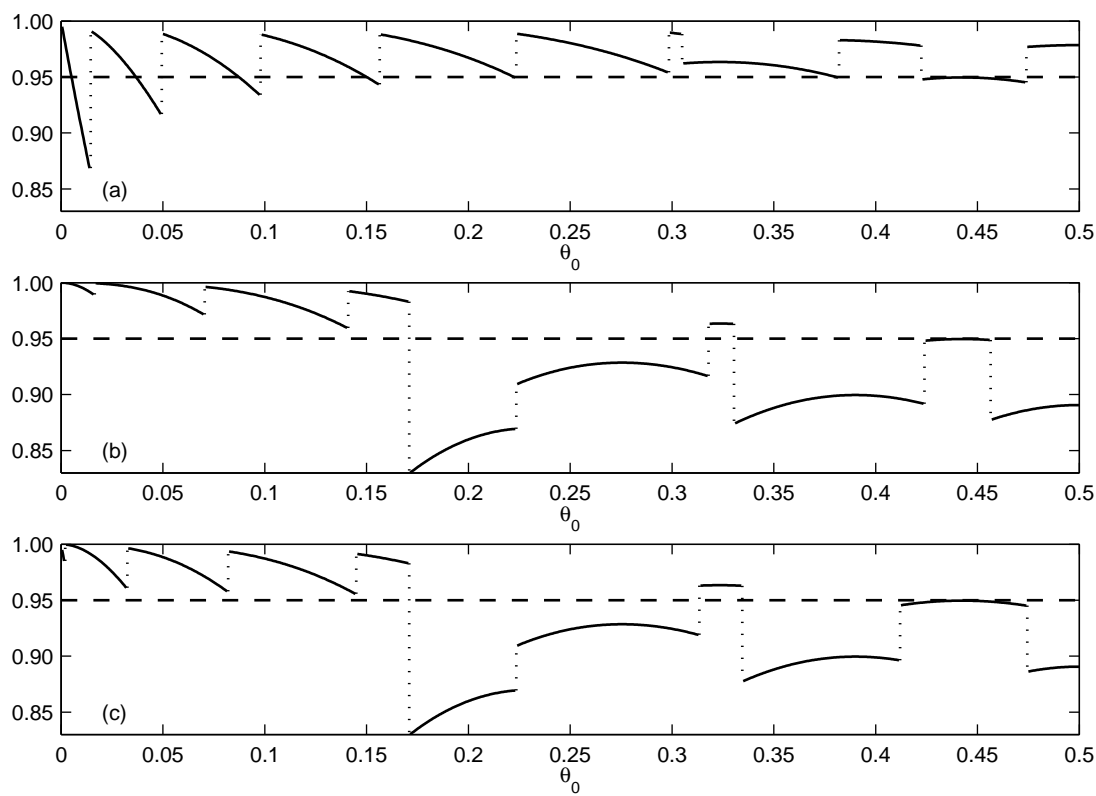}
  \caption{Frequentist coverage of binomial 95\%-credible
    intervals: (a) Predictive criterion 
    estimation and intrinsic estimation with directed
    $\KL(\tilde{\theta}|\theta)$, (b) Intrinsic estimation directed
    $\KL(\theta|\tilde{\theta})$, (c) Intrinsic estimation
    symmetric
    $\min\{\KL(\tilde{\theta}|\theta),\KL(\theta|\tilde{\theta})\}$.}
  \label{fig:berncoverage}
\end{figure}

\begin{figure}[htbp]
  \centering
  \includegraphics{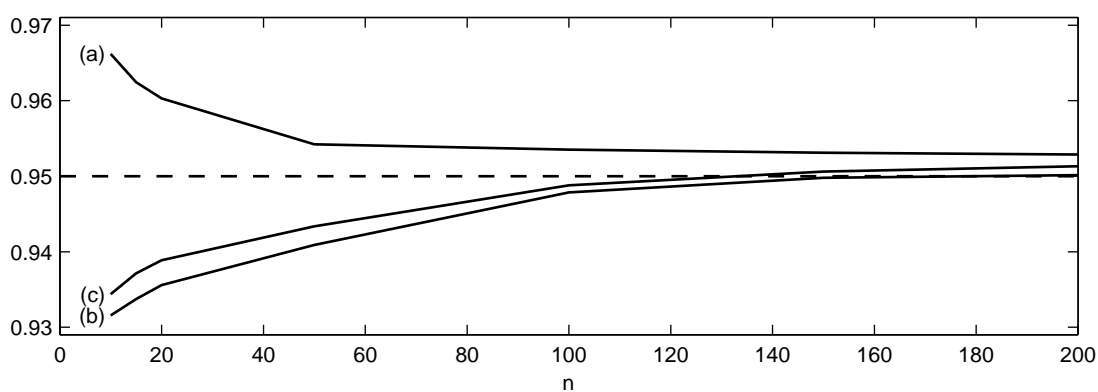}
  \caption{Average coverage over $\theta$ of binomial 95\%-credible
    intervals for different $n$: (a) Predictive criterion
    estimation and intrinsic estimation with directed
    $\KL(\tilde{\theta}|\theta)$, (b) Intrinsic estimation directed
    $\KL(\theta|\tilde{\theta})$, (c) Intrinsic estimation
    symmetric
    $\min\{\KL(\tilde{\theta}|\theta),\KL(\theta|\tilde{\theta})\}$.}
  \label{fig:bernavecoverage}
\end{figure}

Further comparison of the methods is made by comparing the
frequentist coverage of intervals. Since $y$ is discrete, exact
confidence intervals do not exist and coverage is bound to differ
from the target value for most values of true $\theta$.
Figure \ref{fig:berncoverage} shows the frequentist coverage for
different methods.
Predictive criterion estimation and intrinsic estimation with
the directed discrepancy $\KL(\tilde{\theta}|\theta)$ provide exactly
the same result (results shown in the same subplot). 
Intrinsic estimation with the directed discrepancy
$\KL(\theta|\tilde{\theta})$ and intrinsic estimation with the
symmetric discrepancy
$\min\{\KL(\tilde{\theta}|\theta),\KL(\theta|\tilde{\theta})\}$
provide very similar results.
\citet{Bernardo:2005c} speculates that intrinsic credible intervals
with the symmetric
$\min\{\KL(\tilde{\theta}|\theta),\KL(\theta|\tilde{\theta})\}$
might possibly provide the best available solution for this
particular problem, but considering discussion in
\citet{Brown+Cai+Dasgupta:2001a} there is no consensus on which
properties should be preferred. For example, some discussants
prefered intervals whose coverage is allways larger than the target
value. None of the methods discussed in this paper achieve that
completely, although predictive criterion estimation and intrinsic
estimation with the directed discrepancy $\KL(\tilde{\theta}|\theta)$
produce the best result in this sense.
As further illustration, Figure \ref{fig:bernavecoverage} shows the
average coverage for different values of $n$ with different
methods. These plots can be compared to plots of other methods
shown in \citet{Brown+Cai+Dasgupta:2001a}. With all the methods
average coverage converges close to the target value.

\subsection{Exponential model}
\label{sec:example-exponential}

Let $y_{(1:n)}=\{y_1,\ldots,y_n\}$, be a random sample from the
Exponential distribution $\Ex(y|\theta)=\theta e^{-y\theta}$.
Reference prior is $\theta^{-1}$ and corresponding posterior is
Gamma $p(\theta|y_{(1:n)})=\Ga(\theta|n,t)$, where $t=\sum_{i=1}^n
y_i$. As a numerical illustration an example by
\citet{Bernardo:2005b} is used, with $n=10$ and $t=6.08$.

\subsubsection{Predictive criterion estimation}

The predictive distribution for the full model is Gamma-Gamma
\begin{equation}
  \Gg(y|n,t,1)=\frac{\Gamma(n+1)}{\Gamma(n)}\frac{t^n}{(t+y)^{n+1}},
\end{equation}
and the predictive distribution for the restricted model is
exponential
\begin{equation}
  \Ex(y|\tilde{\theta})=\tilde{\theta} e^{-y\tilde{\theta}}.
\end{equation}
The predictive criterion to maximize is
\begin{equation}
  \int \log \Ex(y|\tilde{\theta}) \Gg(y|n,t,1) dy =
  -t\tilde{\theta}/(n-1)+\log\tilde{\theta},
  \label{eq:pred-discr-exp}
\end{equation}
which is maximized by
\begin{equation}
  \hat{\theta}=(n-1)/t.
  \label{eq:pred-est-exp}
\end{equation}

\subsubsection{Intrinsic estimation}

Directed Kullback-Leibler divergences are
\begin{equation}
  \KL(\tilde{\theta}|\theta)=(\tilde{\theta}/\theta)-1-\log(\tilde{\theta}/\theta),
\end{equation}
and
\begin{equation}
  \KL(\theta|\tilde{\theta})=(\theta/\tilde{\theta})-1-\log(\theta/\tilde{\theta}).
\end{equation}
The symmetric intrinsic discrepancy is
\begin{gather}
  \delta(\theta|\tilde{\theta})=
  \begin{cases}
    (\theta/\tilde{\theta})-1-\log(\theta/\tilde{\theta}) & \theta \leq \tilde{\theta} \\
    (\tilde{\theta}/\theta)-1-\log(\tilde{\theta}/\theta) & \theta > \tilde{\theta}.
  \end{cases}
\end{gather}
The expected posterior intrinsic discrepancy for
$\KL(\tilde{\theta}|\theta)$ is
\begin{equation}
  d(\tilde{\theta},y_{(1:n)})=t\tilde{\theta}/(n-1)-\log(\tilde{\theta})-\log(t)-1+\psi(n),
  \label{eq:dir1-intr-discr-exp}
\end{equation}
which is ignoring the term independent of $\tilde{\theta}$
proportional to \eqref{eq:pred-discr-exp}. The expected
discrepancy \eqref{eq:dir1-intr-discr-exp} is minimized by
\begin{equation}
  \hat{\theta}=(n-1)/t,
  \label{eq:dir1-intr-est-exp}
\end{equation}
which is equal to \eqref{eq:pred-est-exp}. The expected posterior
intrinsic discrepancy for $\KL(\theta|\tilde{\theta})$ is
\begin{equation}
  n/(t\tilde{\theta})+\log(\tilde{\theta})+\log(t)-1-\psi(n),
\end{equation}
which is minimized by 
\begin{equation}
  \hat{\theta}=n/t.
  \label{eq:dir2-intr-est-exp}
\end{equation}
The expected posterior intrinsic discrepancy with the symmetric
discrepancy does not have an analytic solution, but can be
evaluated, for example, with quadrature methods.
\citet{Bernardo:2005b} proposes approximation
\begin{equation}
  \hat{\theta}=(n-1/2)/t,
  \label{eq:sym-intr-est-exp}
\end{equation}
which happens to be the average of estimates obtained with directed
Kullback-Leibler divergences.
  
\subsubsection{Results}

Table \ref{tab:results3} shows point estimates and 95\%-credible
intervals of $\theta$ obtained with different methods. Note that,
all credible intervals are in this case exact confidence intervals
\citep[see][]{Bernardo:2005c}. Frequentist confidence intervals are
central intervals. Note that, taking inverse of central interval
endpoints for the scale ($1/\theta$) does not produce exact
confidence interval for the rate $\theta$.
\begin{table}[htbp]
  \centering
  \begin{tabular}[t]{l c c}
    Approach & $\hat{\theta}$ & CI \\ \hline
    Predictive criterion estimation & 1.48 & (0.71,\,2.68) \\
    Intrinsic estimation directed $\KL(\tilde{\theta}|\theta)$ &
    1.48 & (0.71,\,2.68) \\
    Intrinsic estimation directed $\KL(\theta|\tilde{\theta})$ &
    1.64 & (0.89,\,3.58) \\
    Intrinsic estimation symmetric
    $\min\{\KL(\tilde{\theta}|\theta),\KL(\theta|\tilde{\theta})\}$
    & 1.57 & (0.83,\,2.95) \\
    Minimum variance unbiased estimate of rate $\theta$ & 1.48 & (0.79,\,2.81)
    \\
    Inverse of minimum variance unbiased estimate of scale ($1/\theta$) &
    1.64 & (0.96,\,3.43)
  \end{tabular}
  \caption{Point estimates of $\theta$ of exponential model with
    different methods.}
  \label{tab:results3}
\end{table}

Predictive criterion estimation and intrinsic estimation with
the directed discrepancy $\KL(\tilde{\theta}|\theta)$ give exactly same
estimate (see \eqref{eq:pred-est-exp} and
\eqref{eq:dir1-intr-est-exp}).

Intrinsic estimation with directed $\KL(\theta|\tilde{\theta})$
gives higher estimate for the rate $\theta$ and lower estimate for
the scale (mean wait time) $1/\theta$. Estimate for the rate is not
unbiased estimate, but estimate for the scale is the minimum
variance unbiased estimate for the scale.
This is similar to the Normal distribution example where the
intrinsic estimate with directed $\KL(\theta|\tilde{\theta})$ is
the minimum variance unbiased estimate for the scale $\sigma^2$ but
not for inverese of the scale $1/\sigma^2$.
This is due to a fact that that the minimum variance unbiased
estimate is not, but predictive criterion and intrinsic estimates
are invariant under parameterization.

Intrinsic estimation with symmetric divergence gives estimate which
is between the directed estimates, and more specifically the
estimate is approximately the average of the directed estimates.
\citet{Bernardo:2005b} proposes this approximation without noting
its connection to directed estimates.

\subsection{Uniform model}
\label{sec:example-uniform}

The next example illustrates a non-regular problem where the
directed Kullback-Leibler divergences diverge. Let
$y_{(1:n)}=\{y_1,\ldots,y_n\}$, be a random sample from the
Uniform distribution $\Un(y|0,\theta)=\theta^{-1},0<y<\theta$.
Sufficient statistic is $t=\max\{y_1,\ldots,y_n\}$. As a numerical
illustration an example by \citet{Bernardo+Juarez:2003} is used,
with $n=10$ and $t=1.897$. The reference posterior distribution is
the Pareto distribution \citep{Bernardo+Juarez:2003}
\begin{equation}
  p(\theta|y_{(1:n)})=\Pa(\theta|n,t)=n t^n \theta^{-(n+1)}, \quad
  \theta>t.
\end{equation}

\subsubsection{Predictive criterion estimation}

The predictive distribution of the full model is
\begin{gather}
  p(y|y_{(1:n)})=
  \begin{cases}
    \frac{n}{(n+1)t} & \text{if $0\leq y\leq t$}\\
    \frac{n t^n}{(n+1)y^{n+1}} & \text{if $y>t$}.
  \end{cases}
  \label{eq:unif-full-pred}
\end{gather}
The predictive distribution of the restricted model
$\Un(0,\tilde{\theta})$ is 
\begin{gather}
  p(y|\tilde{\theta})=
  \begin{cases}
    \frac{1}{\tilde{\theta}} & \text{if $y\leq \tilde{\theta}$}\\
    0 & \text{otherwise}.
  \end{cases}
  \label{eq:unif-restr-pred}
\end{gather}
The problem is that the logarithmic score based predictive
criterion degenerates since $p(y|\tilde{\theta})=0,
y>\tilde{\theta}$ and the support of $p(y|y_{(1:n)})$ is
$(0,\infty)$.
The logarithmic score measures the information lost when going from
the full model to the restricted model. Going from any finite
probability to zero probability loses an infinite amount of
information, and thus it is not sensible to approximate
\eqref{eq:unif-full-pred} with \eqref{eq:unif-restr-pred} if one
does not want to lose an infinite amount of information about
possible values of $y$ which may be observed.

Furthermore, consider that the predictive distribution would be
used as a prior for some future inference. If the predictive
distribution of the restricted model were used as prior, that prior
would have zero probability for all values larger than $\hat{\theta}$.
Then, observing values large than that would not change the
posterior information. This is related to Cromwell's Rule by
\citet{Lindley:1985a}, which states that one should avoid using
prior probability of 0.

Based on these arguments, it can be further argued that for the
purposes of pure scientific inference and communication the
restriced model in this problem does not make sense. On the other
hand, application specific utilities can still make sense. For
example, if the predictive decision problem would be reformulated
as a guessing contest where the one with the closest guess wins,
the best guess with the corresponding utility function would be the
median of the posterior $\hat{\theta}=2.03$.

\subsubsection{Intrinsic estimation}

Directed Kullback-Leibler divergences are
\begin{gather}
  \KL(\tilde{\theta}|\theta)=
  \begin{cases}
    \log(\tilde{\theta}/\theta) & \theta \leq \tilde{\theta} \\
    \infty & \theta > \tilde{\theta},
  \end{cases}
\end{gather}
and
\begin{gather}
  \KL(\theta|\tilde{\theta})=
  \begin{cases}
    \log(\theta/\tilde{\theta}) & \theta \geq \tilde{\theta} \\
    \infty & \theta < \tilde{\theta}.
  \end{cases}
\end{gather}
Now directed Kullback-Leibler divergences diverge and cannot be
used. \citet{Bernardo+Juarez:2003} solve this problem by using the
symmetric intrinsic discrepancy by taking minimum of above directed
divergences
\begin{gather}
  \delta(\theta|\tilde{\theta})=
  \begin{cases}
    \log(\tilde{\theta}/\theta) & \theta \leq \tilde{\theta} \\
    \log(\theta/\tilde{\theta}) & \theta > \tilde{\theta}.
  \end{cases}
\end{gather}
Note that, not all symmetric intrinsic discrepancies have this
property. For example, the sum or the average of the directed
Kullback-Leibler divergences diverges too.

The expected posterior intrinsic discrepancy is
\begin{equation}
  d(\tilde{\theta},y_{(1:n)})=\int_t^\infty\delta(\tilde{\theta},\theta)\Pa(\theta|n,t)d\theta
= 2\left(\frac{t}{\tilde{\theta}}\right)^n
-n\log\left(\frac{t}{\tilde{\theta}}\right) -1,
\label{eq:unif-exp-pos-int-disc}
\end{equation}
which is minimized by $\tilde{\theta}=2^{1/n}t$, which happens to
be also the median of the reference posterior. In the numerical
example, the intrinsic estimate and the posterior median is
$\hat{\theta}=2.03$, and the minimum variance unbiased estimate is
$\hat{\theta}=(n+1)/nt \approx 2.09$.

The symmetric intrinsic discrepancy $\delta_3$ provides answer in
this problem, but it was achieved by changing the cost of
consequences. It is not entirely clear whether taking the minimum
of two directed discrepancies with different properties is well
justified as a cost for consequences in pure scientific inference
and communication.

\subsection{Hierarchical model}
\label{sec:example-hierarchical}

So far models in the examples have been one-level models. This
section demonstrate a problem with a simple two-level hierarchical
normal model of data $y_{ij}$ with group level effects $\alpha_j$:
\begin{equation}
  \begin{aligned}
    y_{ij} & \sim \N(\mu+\alpha_j,\sigma^2_y), \quad
    i=1,\ldots,n_j,
    \quad j=1,\ldots,J\\
    \alpha_j & \sim \N(0,\sigma_\alpha^2), \quad j=1,\ldots,J.
  \end{aligned}
  \label{eq:hierarchical-normal}
\end{equation}
Data are from the 8-schools example described in \citet[][Ch.
5]{Gelman+Carlin+Stern+Rubin:2003}. Here, the parameters
$\alpha_1,\ldots,\alpha_8$ represent the relative effects of
Scholastic Aptitude Test coaching programs in eight different
schools, and $\sigma_\alpha$ represents the between-school standard
deviations of these effects. \citet[][Ch.
5]{Gelman+Carlin+Stern+Rubin:2003} further simplify the model
\eqref{eq:hierarchical-normal} to
\begin{equation}
  \begin{aligned}
    \bar{y}_{\cdot j} & \sim \N(\mu+\alpha_j,\sigma^2_j), \quad j=1,\ldots,J\\
    \alpha_j & \sim \N(0,\sigma_\alpha^2), \quad j=1,\ldots,J,
  \end{aligned}
\end{equation}
where
\begin{align}
  \bar{y}_{\cdot j} = \frac{1}{n_j}\sum_{i=1}^{n_j} y_{ij}
  \quad \text{and} \quad
  \sigma_j^2 = \frac{\sigma_y^2}{n_j}.
\end{align}
Data are $J=8$, $\bar{y_{(1:n)}}=\{\text{${28}\;{8}\;{-3}\;{7}\;{-1}\;{1}\;{18}\;{12}$}\}$, and
$\boldsymbol{\sigma}=\{15\;10\;16\;11\;9\;11\;10\;18\}$.
Uniform priors for $\mu$ and $\sigma_\alpha$ are used as
recommended by \citet[][Ch. 5]{Gelman+Carlin+Stern+Rubin:2003} and
\citet{Gelman:2006a}.
In this example, terms intrinsic estimation and intrinsic credible
intervals are used liberally, since strictly intrinsic estimation
is defined by \citet{Bernardo+Juarez:2003} only for reference
posteriors obtained using reference priors, and the prior used here
is not a reference prior (see \citet{Yang+Berger:1997} for
reference priors for $\sigma_\alpha$).
Point estimates are computed for the hyperparameters $\mu$ and
$\sigma_\alpha$, that is for the common effect of the coaching and
the between-school standard deviation.

The posterior distribution is not available in analytical form but
it is easily estimated using Monte Carlo methods. To obtain samples
from the posterior distribution the factorized simulation described
in \citet[][p. 137]{Gelman+Carlin+Stern+Rubin:2003} was used. The
posterior can be factorized as
\begin{equation}
  p(\alpha_j,\mu,\sigma_\alpha^2|y_{(1:n)},M_*)=p(\sigma_\alpha^2|y_{(1:n)},M_*)p(\mu|\sigma_\alpha^2,y_{(1:n)},M_*)p(\alpha_j|\mu,\sigma_\alpha^2,y_{(1:n)},M_*).
\end{equation}
Simulation from the marginal distribution
$p(\sigma_\alpha^2|y_{(1:n)},M_*)$ is performed numerically using the
one dimensional grid sampling. Simulation from
$p(\mu|\sigma_\alpha^2,y_{(1:n)},M_*)$ and
$p(\alpha_j|\mu,\sigma_\alpha^2,y_{(1:n)},M_*)$ is easy as they are
Normal distributions.

\subsubsection{Predictive criterion estimation}

There are two choices for prediction. One could predict $y_{\cdot
  j}$ or $y_{\cdot J+1}$. Since $\sigma_j^2$ are given as fixed,
there is no proper model for predicting $\sigma_{J+1}^2$ which
would be required for predicting $y_{\cdot J+1}$. Thus, $y_{\cdot
  j}$ are predicted. Note that, point estimate may be different
depending on the predictive inference one is interested in.
The predictive distribution is
\begin{equation}
  p(\bar{y}|y_{(1:n)},M_*)=\int
  p(\bar{y}|\alpha_j,M_*)p(\sigma_\alpha^2|y_{(1:n)},M_*)p(\mu|\sigma_\alpha^2,y_{(1:n)},M_*)p(\alpha_j|\mu,\sigma_\alpha^2,y_{(1:n)},M_*) d \alpha_j\mu\sigma_\alpha^2.
\end{equation}
Parameters $\alpha_j$ can be integrated out analytically to get the
predictive distribution of the full model in form
\begin{equation}
  p(\bar{y}|y_{(1:n)},M_*)=\int
  p(\bar{y}|\mu,\sigma_\alpha^2,M_*)p(\sigma_\alpha^2|y_{(1:n)},M_*)p(\mu|\sigma_\alpha^2,y_{(1:n)},M_*) d \mu \sigma_\alpha^2,
\end{equation}
which is a mixture of normal distributions. By integrating out
parameters $\alpha_j$, the predictive distribution of the
restricted model is
\begin{equation}
  p(\bar{y}|y_{(1:n)},M_R)=p(\bar{y}|\tilde{\mu},\tilde{\sigma}_\alpha^2),
\end{equation}
which is a normal distribution with parameters whose values can be
computed given data and the values of $\tilde{\mu}$ and
$\tilde{\sigma}_\alpha^2$. The value of the predictive criterion
given $\tilde{\mu}$ and $\tilde{\sigma}_\alpha^2$ is easily
computed using combination of Monte Carlo sampling of $\mu$ and
$\sigma_\alpha^2$ and quadrature integration for the rest.

\subsubsection{Intrinsic estimation}

Intrinsic estimation is not directly applicable for hierarchical
models. The intrinsic discrepancy \eqref{eq:intrinsic-discrepancy}
does not have reference to hyperparameters as the discrepancy is
computed using only likelihood part of the model. To overcome this
problem, it is possible to integrate over the parameters $\alpha_j$
and use the predictive distribution conditional on the
hyperparameters
\begin{equation}
  p(\bar{y}|\mu,\sigma_\alpha^2),
\end{equation}
which is a normal distribution with parameters whose values can be
computed given data and values of $\mu$ and $\sigma_\alpha^2$.
Kullback-Leibler divergence equations are given in section
\ref{sec:example-normal} and the expected discrepancy is computed
over the posterior distribution of $\mu$ and $\sigma_\alpha^2$. The
expectation is easily estimated using Monte Carlo simulation.

\subsubsection{Results}

Table \ref{tab:results5} shows point estimates of $\mu$ and
$\sigma_\alpha$ obtained with different methods. Note that,
classical unbiased estimate based on analysis of variance fails by
estimating $\sigma_\alpha^2$ to be negative \citep[see][Ch.
5]{Gelman+Carlin+Stern+Rubin:2003}.
\begin{table}[htbp]
  \centering
  \begin{tabular}[t]{l c c}
    Approach & $\mu^*$ & $\sigma_\alpha^*$ \\ \hline
    Predictive criterion estimation & 8.0 & 7.3 \\
    Intrinsic estimation directed $\KL(\tilde{\theta}|\theta)$ & 8.0 & 7.3 \\
    Intrinsic estimation directed $\KL(\theta|\tilde{\theta})$ & 7.9 & 5.7 \\
    Intrinsic estimation symmetric $\min\{\KL(\tilde{\theta}|\theta),\KL(\theta|\tilde{\theta})\}$ & 7.9 & 6.7 
  \end{tabular}
  \caption{Point estimate of $\mu$ and $\sigma_\alpha$ of two-level
    normal model with different methods.}
  \label{tab:results5}
\end{table}
As in other examples, predictive criterion estimation and intrinsic
estimation with directed divergence $\KL(\tilde{\theta}|\theta)$
give the same estimate.
Intrinsic estimation with directed $\KL(\theta|\tilde{\theta}$)
gives lower estimate for $\sigma_\alpha$ which is in line with
results for one-level normal model in section
\ref{sec:example-normal}. The predictive approach includes the
uncertainty form the full model predictive inference by giving more
pessimistic point estimates.
As in other examples, intrinsic estimation with the symmetric
divergence gives an estimate which is between directed estimates,
although it is not as close to the average of the directed
estimates as in the other examples.

\section{Conclusion}
\label{sec:conclusion}

The selection of utility function should not be arbitrary. The
purpose of this paper was to examine properties of the intrinsic
loss function used in intrinsic estimation and credible regions by
\citet{Bernardo+Juarez:2003} and \citet{Bernardo:2005c}, and
compare them to the predictive criterion. Intrinsic estimation
tries to provide a reference point estimate without an explicit
application specific utility. If an application specific loss
function were available, naturally it should be used instead of
intrinsic loss function. The problem is that when application
specific utility is not available, there is no unambiguous choice
of intrinsic loss function.

Intrinsic estimation with the directed discrepancy
$\KL(\tilde{\theta}|\theta)$ is strongly related to the predictive
criterion approach suitable for finding a restricted model with the
best predictive performance. This can also be interpreted as
finding a restricted model with as similar predictive density as
possible as the predictive density of the full model. This approach
is useful if the restricted model is going to be used for further
predictive inference. If there are no nuisance parameters the
methods are equal, and with presence of nuisance parameters methods
produce similar estimates. The predictive criterion is generally
easier to compute for arbitrary models (e.g., with help of MCMC)
and it's interpretation is more obvious.

Intrinsic estimation with the directed discrepancy
$\KL(\theta|\tilde{\theta})$ is related to the frequentist approach
assuming there exists a true fixed value, which generated the data,
and the goal is to find the best estimate for that value. Specially
in three regular problems, estimates for the location and scale
parameters were also unbiased minimun variance estimates. This
approach might be useful if good frequency properties are desired,
although further research is required.

Intrinsic estimation with the symmetric discrepancy
$\min\{\KL(\tilde{\theta}|\theta),\KL(\theta|\tilde{\theta})\}$ is
compromise between the estimation with the directed discrepancies.
The use of the symmetric discrepancy seems to be not convincingly
justified.
Estimates are not as good for further predictions as estimates with
the directed discrepancy $\KL(\tilde{\theta}|\theta)$ nor do they
have as good frequency properties as estimates with the directed
discrepancy $\KL(\theta|\tilde{\theta})$.
Even in pure scientific inference and communication, one should
probably know, which one of these properties is preferred for
estimates.

\section{Discussion}
\label{sec:discussion}

The results in this paper have also relevance to hypothesis testing
and model selection.  Intrinsic discrepancy based hypothesis
testing was proposed by \citet{Bernardo+Rueda:2002} and further
advocated by \citet{Bernardo:2005a}. \citet{Bernardo+Rueda:2002}
define hypothesis testing as a model selection problem where the
full model $M_*$ is tentatively accepted, and it is desired to test
whether a restricted model $M_R$ is compatible with observed data.
Since in the Bayesian approach we should integrate over the model
uncertainty, the full model $M_*$ should automatically include
then alternative models as submodels, and all model selection
matches with this defintion of hypothesis testing.

In point estimation, the restricted model $M_R$ obtained by
minimizing the expected discrepancy was automatically accepted.
Regardless of whether $M_R$ is obtained by point estimation or by
fixing $\theta$ to some value with special meaning in the model
(e.g. zero), in hypothesis testing also the value of the expected
discrepancy is considered. The problem is how to calibrate the
discrepancy, that is, how large exepected discrepancy can be
accepted without rejecting the hypothesis.
\citet{Bernardo:1999a} and \citet{Bernardo+Rueda:2002} note the
connection of the expected discrepancy to the log-likelihood ratio
and propose that scale used for the log-likelihood ratio could be
used.
\citet{Bernardo:1999a} used the directed discrepancy
$\KL(\tilde{\theta}|\theta)$ and \citet{Bernardo+Rueda:2002} the
symmetric discrepancy
$\min\{\KL(\tilde{\theta}|\theta),\KL(\theta|\tilde{\theta})\}$.
With same argumentation the log-likelihood ratio scale would be
suitable also for the directed discrepancy
$\KL(\tilde{\theta}|\theta)$ and the predictive criterion.

Three intrinsic discrepancies considered in this paper all give
different expected discrepancies, with different minima and shapes.
It seems plausible that propeties in hypothesis testing would be
similar to properties in point estimation, although this requires
further research.

\section*{Acknowledgments}
\addcontentsline{toc}{section}{Acknowledgments}
  
The author would like to thank Jouko Lampinen, Ilkka Kalliomäki,
and Harri Valpola for helpful comments and suggestions.

\bibliography{brc}

\begin{thebibliography}{}

\bibitem[Bayarri and Berger, 2004]{Bayarri+Berger:2004a}
Bayarri, M.~J. and Berger, J.~O. (2004).
\newblock The interplay of {Bayesian} and frequentist analysis.
\newblock {\em Statistical Sceience}, 19(1):58--80.

\bibitem[Bernardo, 1979a]{Bernardo:1979}
Bernardo, J.~M. (1979a).
\newblock Expected information as expected utility.
\newblock {\em Annals of Statistics}, 7(3):686--690.

\bibitem[Bernardo, 1979b]{Bernardo:1979b}
Bernardo, J.~M. (1979b).
\newblock Reference posterior distributions for {Bayesian} inference.
\newblock {\em Journal of the Royal Statistical Society. Series B
  (Methodological)}, 41(2):113--147.

\bibitem[Bernardo, 1999]{Bernardo:1999a}
Bernardo, J.~M. (1999).
\newblock Nested hypothesis testing: The {Bayesian} reference criterion.
\newblock In Bernardo, J.~M., Berger, J.~O., and Dawid, A.~P., editors, {\em
  Bayesian Statistics 6}, pp. 101--130. Oxford University Press.

\bibitem[Bernardo, 2005a]{Bernardo:2005b}
Bernardo, J.~M. (2005a).
\newblock An integrated mathematical statistics primer: Objective {Bayesian}
  construction, frequentist evaluation.
\newblock {\em ISI Bulletin}.
\newblock In press.

\bibitem[Bernardo, 2005b]{Bernardo:2005c}
Bernardo, J.~M. (2005b).
\newblock Intrinsic credible regions: An objective bayesian approach to
  interval estimation.
\newblock {\em Test}, 14(2).
\newblock 317--384.

\bibitem[Bernardo, 2005c]{Bernardo:2005a}
Bernardo, J.~M. (2005c).
\newblock Reference analysis.
\newblock In Dey, D. and Rao, C.~R., editors, {\em Handbook of Statistics},
  volume~25. Elsevier.
\newblock 17--90.

\bibitem[Bernardo and Juárez, 2003]{Bernardo+Juarez:2003}
Bernardo, J.~M. and Juárez, M.~A. (2003).
\newblock Intrinsic estimation.
\newblock In Bernardo, J.~M., Bayarri, M.~J., Berger, J.~O., Dawid, A.~P.,
  Heckerman, D., Smith, A. F.~M., and West, M., editors, {\em Bayesian
  Statistics 7}, pp. 456--476. Oxford University Press.

\bibitem[Bernardo and Rueda, 2002]{Bernardo+Rueda:2002}
Bernardo, J.~M. and Rueda, R. (2002).
\newblock Bayesian hypothesis testing: a reference approach.
\newblock {\em International Statistical Review}, 70(3):351--372.

\bibitem[Bernardo and Smith, 1994]{Bernardo+Smith:1994}
Bernardo, J.~M. and Smith, A. F.~M. (1994).
\newblock {\em Bayesian Theory}.
\newblock John Wiley \& Sons.

\bibitem[Brown et~al., 2001]{Brown+Cai+Dasgupta:2001a}
Brown, L.~D., Cai, T.~T., and DasGupta, A. (2001).
\newblock Interval estimation for a binomial proportion (with discussion).
\newblock {\em Statistical Science}, 16(2):101--133.

\bibitem[Catalina et~al.,
  2021]{Catalina+Buerkener+Vehtari:2021:latent_projpred}
Catalina, A., B{\"u}rkner, P., and Vehtari, A. (2021).
\newblock Latent space projection predictive inference.
\newblock {\em arXiv preprint arXiv:2109.04702}.

\bibitem[Catalina et~al., 2020]{Catalina+Buerkner+Vehtari:2020:GAMM_projpred}
Catalina, A., B{\"u}rkner, P.-C., and Vehtari, A. (2020).
\newblock Projection predictive inference for generalized linear and additive
  multilevel models.
\newblock {\em arXiv preprint arXiv:2010.06994}.

\bibitem[Dey et~al., 1994]{Dey+Gelfand+Swartz+Vlachos:1994}
Dey, D.~K., Gelfand, A.~E., Swartz, T.~B., and Vlachos, P.~K. (1994).
\newblock A simulation-intensive approach for checking hierarchical models.
\newblock Technical Report tr9529, Department of Statistics, University of
  Connecticut.

\bibitem[Dupuis and Robert, 2003]{Dupuis+Robert:2003}
Dupuis, J.~A. and Robert, C.~P. (2003).
\newblock Variable selection in qualitative models via an entropic explanatory
  power.
\newblock {\em Journal of Statistical Planning and Inference}, 111:77--94.

\bibitem[Gelfand and Ghosh, 1998]{Gelfand+Ghosh:1998}
Gelfand, A.~E. and Ghosh, S.~K. (1998).
\newblock Model choice: A minimum posterior predictive loss approach.
\newblock {\em Biometrika}, 85:1--11.

\bibitem[Gelman, 2006]{Gelman:2006a}
Gelman, A. (2006).
\newblock Prior distributions for variance parameters in hierarchical models
  (comment on article by browne and draper).
\newblock {\em Bayesian Analysis}, 1(3):515--534.

\bibitem[Gelman et~al., 2013]{Gelman+etal+BDA3:2013}
Gelman, A., Carlin, J.~B., Stern, H.~S., Dunson, D.~B., Vehtari, A., and Rubin,
  D.~B. (2013).
\newblock {\em Bayesian Data Analysis}.
\newblock Chapman \& Hall/CRC, third edition.

\bibitem[Gelman et~al., 2003]{Gelman+Carlin+Stern+Rubin:2003}
Gelman, A., Carlin, J.~B., Stern, H.~S., and Rubin, D.~R. (2003).
\newblock {\em Bayesian Data Analysis}.
\newblock Chapman \& Hall, 2nd edition.

\bibitem[Gelman et~al., 1996]{Gelman+Meng+Stern:1996}
Gelman, A., Meng, X.-L., and Stern, H. (1996).
\newblock Posterior predictive assessment of model fitness via realized
  discrepancies (with discussion).
\newblock {\em Statistica Sinica}, 6(4):733--807.

\bibitem[Goutis and Robert, 1998]{Goutis+Robert:1998a}
Goutis, C. and Robert, C.~P. (1998).
\newblock Model choice in generalised linear models: A {Bayesian} approach via
  {Kullback}-{Leibler} projections.
\newblock {\em Biometrika}, 85(1):29--37.

\bibitem[Guti{\'e}rrez-Pe{\~n}a and Walker, 2001]{Gutierrez-Pena+Walker:2001}
Guti{\'e}rrez-Pe{\~n}a, E. and Walker, S.~G. (2001).
\newblock A {Bayesian} predictive approach to model selection.
\newblock {\em Journal of Statistical Planning and Inference},
  93(1--2):259--276.

\bibitem[Laud and Ibrahim, 1995]{Laud+Ibrahim:1995}
Laud, P. and Ibrahim, J. (1995).
\newblock Predictive model selection.
\newblock {\em Journal of the Royal Statistical Society. Series B
  (Methodological)}, 57:247--262.

\bibitem[Lindley, 1968]{Lindley:1968a}
Lindley, D.~V. (1968).
\newblock The choice of variables in multiple regression.
\newblock {\em Journal of the Royal Statistical Society. Series B
  (Methodological)}, 30(1):31--66.

\bibitem[Lindley, 1985]{Lindley:1985a}
Lindley, D.~V. (1985).
\newblock {\em Making decisions}.
\newblock John Wiley \& Sons, 2nd edition edition.

\bibitem[{O'Hagan} and Forster, 2004]{OHagan+Forster:2004}
{O'Hagan}, A. and Forster, J. (2004).
\newblock {\em Bayesian Inference}, volume~2B of {\em Kendalls´s Advanced
  Theory of Statistics}.
\newblock Arnold, 2nd edition.

\bibitem[Piironen et~al., 2020]{Piironen+Paasiniemi+Vehtari:2020:projpred}
Piironen, J., Paasiniemi, M., and Vehtari, A. (2020).
\newblock Projective inference in high-dimensional problems: Prediction and
  feature selection.
\newblock {\em Electronic Journal of Statistics}, 14(1):2155--2197.

\bibitem[Press, 2003]{Press:2003}
Press, S.~J. (2003).
\newblock {\em Subjective and Objective Bayesian Statistics: Principles,
  Models, and Applications}.
\newblock John Wiley \& Sons.

\bibitem[Robert, 1996]{Robert:1996b}
Robert, C.~P. (1996).
\newblock Intrinsic losses.
\newblock {\em Theory and decision}, 40(2):191--214.

\bibitem[Robert, 2001]{Robert:2001}
Robert, C.~P. (2001).
\newblock {\em The Bayesian Choice: from Decision-Theoretic Motivations to
  Computational Implementation}.
\newblock Springer, 2nd edition.

\bibitem[{San Martini} and Spezzaferri, 1984]{SanMartini+Spezzaferri:1984}
{San Martini}, A. and Spezzaferri, F. (1984).
\newblock A predictive model selection criterion.
\newblock {\em Journal of the Royal Statistical Society. Series B
  (Methodological)}, 46(2):296--303.

\bibitem[Trevisani and Gelfand, 2003]{Trevisani+Gelfand:2003a}
Trevisani, M. and Gelfand, A.~E. (2003).
\newblock Inequalities between expected marginal log likelihoods with
  implications for likelihood-based model comparison.
\newblock {\em The Canadian Journal of Statistics}, 31(3):239--250.

\bibitem[Vehtari and Ojanen, 2012]{Vehtari+Ojanen:2012}
Vehtari, A. and Ojanen, J. (2012).
\newblock A survey of {Bayesian} predictive methods for model assessment,
  selection and comparison.
\newblock {\em Statistics Surveys}, 6:142 -- 228.

\bibitem[Yang and Berger, 1997]{Yang+Berger:1997}
Yang, R. and Berger, J.~O. (1997).
\newblock A catalog of noninformative priors.
\newblock ISDS Discussion Paper 97-42, Institute of Statistics and Decision
  Sciences, Duke University.

\end{thebibliography}

\end{document}